\documentclass[aps,pra, onecolumn,14pt,a4paper,showpacs]{revtex4}

\usepackage{amssymb,amsmath,amsfonts,graphicx}
\usepackage{bm}

\begin{document}

\title{Optical solitons as quantum objects}

\author{ Yves Pomeau }

\affiliation{Laboratoire de physique statistique de l'Ecole normale
sup\'erieure, \\ \small 24 Rue Lhomond, 75231 Paris Cedex 05,
France.}

\author{Martine Le Berre}

\affiliation{Laboratoire de photophysique mol\'eculaire, Bat.210,
91405 Orsay, France}

\date{\today }

\begin{abstract}

The intensity of classical bright solitons propagating in  linearly
coupled identical fibers can be distributed either in a stable
symmetric state at strong coupling or in a stable asymmetric state
if the coupling is small enough. In the first case, if the initial
state is  not the equilibrium state, the intensity may switch
periodically from fiber to fiber, while in  the second case the
a-symmetrical state remains forever, with most of its energy in
either fiber. The latter situation makes a state of propagation with
two exactly reciprocal realizations. In the quantum case, such a
situation does not exist as an eigenstate because of the quantum
tunneling between the two fibers. Such a tunneling is a purely
quantum phenomenon which does not not exist in the classical
theory. We estimate the rate of tunneling by quantizing a simplified
dynamics derived from the original Lagrangian equations with test
functions. This tunneling could be within reach of the experiments,
particularly if the quantum coherence of the soliton can be
maintained over a sufficient amount of time.
\bigskip

{\emph{Lead Paragraph}}

\textbf{Usually solitons in optical fibers are assumed to be
classical (= non quantum) objects because they are made of a large
number of photons. Nevertheless there exist quantum effects without
classical counterpart, like the tunneling under a potential barrier.
We investigate one possible realization of such a quantum tunneling
with solitons as basic entities. Specifically, we consider a soliton
propagating in two linearly coupled fibers that are assumed
identical. It has been known for some time that, at small enough
coupling, asymmetric solitons only can propagate and be stable. The
amplitude of such asymmetric solitons is predominantly in either
fiber and remains there forever classically. This makes, for a given
energy, two possible steady states exactly symmetrical with respect
to each other under permutation of the two fibers. In the quantum
version of the same problem, the two solitons merge into a single
quantum state sharing a quantum amplitude spread between the two
fibers, because of the possibility of {\emph{quantum}} tunneling
from one fiber to the other. We study this problem thanks to a
reduced set of equations derived from the full set of coupled
nonlinear PDE's by choosing convenient trial functions for the
classical soliton dynamics. Thanks to this choice, the bifurcation
pattern of the soliton solution in the coupled fibers is well
reproduced. Because the trial set generates dynamical equations with
a Lagrange structure, this Lagrangian system is relatively easy to
quantize. To obtain the quantum amplitude of transmission by
tunneling under the barrier, one replaces the original Hamiltonian
system by its Euclidean counterpart. Orders of magnitude relevant
for a possible physical application are given.}

\end{abstract}

\maketitle

\section{Introduction}

Generally speaking a soliton is a localized solution of an equation
for the propagation of a field envelope. It stays localized under
the opposite effects of linear dispersion tending to spread the wave
and of nonlinearity making the wave steeper. We make one more step
by considering this soliton as a `true' particle, that is by seeing
it as a classical object that should be ultimately quantized to keep
the consistency of our view of the physical world. This is of course
not a new idea, see for instance the review \cite{faddeev} on the
quantization of various nonlinear equations for classical fields.
Quantum effects are irrelevant for macroscopic phenomena like
solitary waves in a water channel. However, there is an instance of
solitonic physics where quantization could bring significant new
effects, namely the propagation of optical solitons in fibers: there
the amplitude of the wave may be small enough to yield solitons with
not too large action, measured in units of Planck's constant
$\frac{h}{2\pi}$. If the action is much larger than this quantum
unit one is in the classical regime, many trajectories contribute to
the saddle point of the Feynman integral \cite{Feynman} and quantum
interferences between coherent quantum states become practically
impossible. Conversely, if the action is not too large compared to
this quantum unit, one could observe quantum phenomena as tunneling
and interferences. Moreover, even if the quantum state has a
coherence time shorter than the tunneling time, there is still
quantum tunneling, but at a reduced rate because the build-up of the
state on the other side of the barrier is slowed down \cite{PomSim}.
 Below
we show that quantum tunneling of a soliton may occur between two
weakly coupled fibers, and we discuss the possibility of quantum
interference between the two states carried by each fiber. The
starting point of our study is the well established result that
{\emph{classically}} a soliton injected in a given fiber cannot
switch to the other fiber, when the coupling is less than a certain
critical value. In this range  the soliton evolves towards the
stable a-symmetric solution having its energy predominantly in the
initial fiber \cite{Akh}. There exist another stable solution which
is obtained by permutation of the two fibers. At small coupling the
two a-symmetrical states are separated by a finite barrier that
cannot be crossed classically. We predict that this may be wrong in
practice because of quantum tunneling. Our derivation is based on
the calculation of the tunneling probability which writes in the WKB
approximation, as $T=exp(-2S/\hbar)$, where $S$ is the physical
action associated to tunneling \cite{griffiths}. The quantity
$S/\hbar$ may be small, even for a pulse with a large number of
photons, because all dynamical phenomena we consider, like the
balance between the nonlinearity, the group velocity dispersion and
the coupling, imply small perturbations to the dominant effect
resulting from the linear dispersionless terms of Maxwell's
equations. The perturbations we consider are as small as the
nonlinear term $n_{2}I$ with respect to the dominant term $n_{0}$,
in the expansion of the refractive index $n = n_{0}+n_{2}I$ for the
Kerr medium of the fiber, $I$ being the optical intensity. In this
sense a soliton is a bound state of photons: photons are attracted
to each other by the focusing nonlinearity. The soliton resembles
the atom of a heavy element which is a quantum object made of many
electrons and nucleons. Atomic physics has also to do with energies
much smaller than the rest energy of the particles (electrons and
nucleons) making the atom. Another idea of atomic (and quantum!)
physics is relevant for our goal: If the atom remains in its ground
state, it is able to make interferences with a wavenumber depending
on its mass and velocity only, independently of the details of the
state of its electrons and nucleons. In our study we assume the
absorption and change of frequency of the photons by inelastic
and/or Raman scattering to be negligible, and discuss the role of
these effects in interference experiments in the last section.

Let us outline the organization of this paper. In the Section
\ref{sec:classicalprop} we introduce the classical model of
propagation of solitons in fibers. First we deal with the single
fiber, then with the two coupled fibers. There is nothing new here and
we focus on what is relevant for us, namely the bifurcation diagram
as a function of the coupling. For strong coupling the stable solution
is symmetric (with intensity equally shared between the two fibers). Below a critical coupling, the classical prediction is that the
stable soliton is a-symmetric, with a high amplitude in one fiber and
a small amplitude in the other, as said above. The details of the
transition are a bit complex because it is subcritical. This has
been studied \cite{Akh} by direct
numerical solution of the coupled PDE's describing this problem
(equations (\ref{eq:coupled1}) and (\ref{eq:coupled2}) below).
However interesting it is, this model presents some difficulties for
our quantization problem. Therefore, in the next section
\ref{sec:classicalproptrial} we outline another approach to the same
problem, namely we use the Lagrange formalism to compute approximate
solutions with trial functions (instead of the full unknown
solution). Thanks to an approriate choice of these functions, the
pattern of bifurcations of the asymmetric to symmetric solitons,
known from the direct numerical simulations, is well
recovered. We use the same trial functions as Malomed et al.
\cite{Malomed} who discuss very thoroughly the general issue in a
paper that we recommend to the interested reader.

As explained in section \ref{sec:quantization}, this `trial
dynamics' is used then to quantize the system. This is of course not
exact, but requires far less formalism than the full quantization of
the two coupled nonlinear field equations. Thanks to this method,
one can use standard results and methods of quantum mechanics for
systems with a few degrees of freedom (as opposed to field
theories). In particular, we can compute the trajectory under the
potential barrier, found by multiplying the propagation variable z,
and the Hamilton-Jacobi action $S$ by $i$. This yields a well
defined problem of Hamiltonian mechanics, called sometimes the
Euclidean version of the initial problem. The tunneling factor is
derived from the action of the heteroclinic trajectory joining the
two equilibria: a stable equilibrium in the original Hamiltonian
system remains an equilibrium in the Euclidean one, but it becomes
unstable there. It turns out that the tunneling probability depends
algebraically on the coupling between the two fibers. This is a
significant remark for possible applications, because it yields a
much smoother dependence with respect to the coupling than the usual
exponentially small tunneling amplitudes.

The last section summarizes the main results of this paper,
discusses the possibility of interferences and presents some ideas
for possible applications. Quantitative predictions rest on the
rather complex problem of turning back from the dimensionless
equations used throughout this work to quantities with a physical
dimension, something done in the Appendix.

\section{Classical propagation of solitons in coupled fibers: the
general model} \label{sec:classicalprop}

The mathematical model for the dissipationless propagation of
optical solitons in one fiber is the {\emph{classical}} (= non
quantum) nonlinear Schr\"odinger equation:
\begin{equation}
i\frac{\partial E}{\partial z} + \alpha \frac{\partial^{2}
E}{\partial t^{2}} +\beta|E|^2 E = 0 \mathrm{.} \label{eq:NLS}
\end{equation}
Even though this equation is called nonlinear Schr\"odinger (NLS),
it does {\emph{not}} mean at all that it makes a quantum system. It
only resembles the usual Schr\"odinger equation, but it describes a
purely classical field, exactly as Maxwell's equations do for an EM field.
This equation is written with real coefficients
$\alpha$ and $\beta$ carrying a physical dimension to make possible
the discussion (see Appendix) of the order of magnitude of the
physical effects to be expected. The field $E$ is the complex
amplitude of the electric field in the wave. We take it as a scalar,
although polarization effects could be brought into the picture in
principle. This equation is derived in the Fresnel approximation,
assuming that the changes of amplitude along the fiber are much
slower and on much longer scales than the oscillations of the
optical field itself and it is also written in the frame of
reference moving with the speed of the envelop of the wave, where
the position variable is $z$. For $\alpha$ and $\beta$ real, this
equation has a Lagrange-like structure. It cancels the first order
variation of the `action'
\begin{equation}
{\mathcal{S}} = \int dz \int {\mathrm{d}}t \left[\frac{i}{2}\left(
\overline{E}\frac{\partial E}{\partial z} - E\frac{\partial
\overline{E}}{\partial z}\right) - \alpha |\frac{\partial
E}{\partial t}|^2 + \frac{\beta}{2}|E|^4\right] \mathrm{.}
\label{eq:NLSLagr}
\end{equation}
In this equation $\overline{E}$ is the complex conjugate of $E$. The
writing of the action in equation (\ref{eq:NLSLagr}) brings in an
important problem, because it is not `the' physical action. Such a
physical action has to have the dimension of the product of an
energy and of a time. Therefore,  the action written in equation
(\ref{eq:NLSLagr}) cannot be an action from the point of view of
physical dimensions. The physical action of the EM field is
proportional to $ {\mathcal{S}}$, its derivation is postponed to the
Appendix. An overall constant multiplying factor does not
change the Euler-Lagrange equations, but it is crucial when
quantizing this system because this relies on a comparison between
the action and $\hbar$, two quantities with the same physical
dimension.

By rescaling $E\rightarrow u \beta^{-1/2}$ (assuming $\beta$
positive to be in the focusing case where solitons exist), and $
t\rightarrow t(2\alpha)^{1/2}$, one obtains the dimensionless
nonlinear Shr\"odinger equation:
\begin{equation}
i\frac{\partial u}{\partial z} + \frac{1}{2} \frac{\partial^{2} u}{\partial
t^{2}} +|u|^2 u = 0 \mathrm{.} \label{eq:NLSdim}
\end{equation}

This equation has a number of interesting symmetries. In addition to
the Galilean invariance (if $u(z,t)$ is a solution, then $u(z,
t-z/C)e^{\frac{i}{C}(t-\frac{z}{2C})}$ is also a solution),  it has
a dilation symmmetry: if $u(z,t)$ is a solution and $\mu$ an
arbitrary real number, then $\mu\,u(z\mu^{2}, t\mu)$ is also a
solution. It has a two parameters family of soliton solutions:
\begin{equation}
u_{0} = \frac{\nu e^{i\varphi}}{\cosh(\nu t)} = \nu e^{i\varphi}
\mathrm{sech}(\nu t) \mathrm{.} \label{eq:soliton}
\end{equation}
In this solution, $\nu$ is any real number and the
phase $\varphi$ is $\varphi = \frac{1}{2}\nu^{2} z +\varphi_{0}$,
with $\varphi_{0}$ arbitrary constant phase.

Note that  $z$ plays here the same role as the time in the usual
Schr\"odinger equation. Among the conserved quantities associated to
any solution of the NLS equation, let us write the "energy"

\begin{equation}
{\mathcal{H}} = \frac{1}{2} \int {\mathrm{d}}t \left[|\frac{\partial u}{\partial
t}|^2 -  |u|^4\right] \mathrm{.} \label{eq:energie}
\end{equation}

 Suppose now that, instead of a
single optical fiber, we have two identical coupled fibers, and that
the coupling is linear and preserves the symmetry between the
fibers. The propagation of solitons in this system has been studied
in the last fifteen years \cite{Akh}, \cite{Malomed}-\cite{smyth}.

To describe the two coupled fibers supporting solitons
we introduce two focusing NLS equations, written in a
dimensionless form:
\begin{equation}
i\frac{\partial u}{\partial z} + \frac{1}{2}\frac{\partial^{2}
u}{\partial t^{2}}+|u|^2 u = -\kappa v \mathrm{,}\label{eq:coupled1}
\end{equation}
and
\begin{equation}
i\frac{\partial v}{\partial z} + \frac{1}{2}\frac{\partial^{2}
v}{\partial t^{2}}+|v|^2 v = -\kappa u \mathrm{,}
\label{eq:coupled2}
\end{equation}
where $\kappa$ is the strength of the linear coupling, and we define
the "mass"
\begin{equation}
Q=\int (\left| u\right| ^{2}+\left| v\right| ^{2})dt \mathrm{.}
\label{eq:mass}
\end{equation}
which is a constant of motion. Consider solutions of the form
$u(z,t) = U(t)e^{iqz}$ and $v(z,t) = V(t)e^{iqz}$. Because of the
common phase factor $e^{iqz}$ the $z$-dependence cancels out and the
two functions $U(t)$ and $V(t)$ are solutions of the two coupled
ordinary differential equations:

\begin{equation}
\left \{ \begin{array}{l}

\ -q U + \frac{1}{2}\frac{{\mathrm{d}}^2
U}{{\mathrm{d}} t^{2}}+|U|^2 U = -\kappa V \\
\ -q V + \frac{1}{2}\frac{{\mathrm{d}}^2 V}{{\mathrm{d}}
t^{2}}+|V|^2 V = -\kappa U \mathrm{.} \label{eq:coupledODE2}
\end{array}
\right.
\end{equation}

For the solution to decrease to zero when $t$ tends to plus or minus
infinity one must have $q>0$. Furthermore the sign of $\kappa$ can
be changed by changing $U$ into $-U$ for instance and keeping $V$
the same.  We choose $\kappa$ positive that corresponds to in-phase
stationary solutions ($U,V$), the out of phase ones being unstable
\cite{Akh}.

This set of equations has been studied numerically and analytically
\cite{Akh}. An exact calculation shows that the symmetric solution
$U=V$ always exists and is linearly stable in the range
$\frac{Q}{\sqrt{\kappa }}\leq \frac{8}{\sqrt{3}}$. For higher values
of this ratio, the symmetric solution looses its stability and an
a-symmetric solution branches off. While it is not explicitely
mentioned in \cite{Akh}, the subcritical character of the
bifurcation can be deduced from Fig.11 of the paper by Akhmediev and
Soto-Crespo(1994) when using $\frac{Q}{\sqrt{\kappa }}$ as control
parameter and $ \frac{q}{\kappa}$ as order parameter, i.e. by
rotating the figure.

Consequently no stable and weakly asymmetric solutions branches off
the unstable symmetric soliton for $\kappa$ slighty smaller than the
onset of linear stability, although an unstable asymmetric solution
branches off at values of $\kappa$ slightly larger than the critical
one. Furthermore a branch of stable asymmetric solitons goes
continuously from $\kappa =0$ to a finite coupling, slightly larger
than the value of linear instability of the symmetric soliton. The
stable asymmetric soliton disappears by a saddle-node bifurcation
for a value of the coupling that is, by a numerical coincidence,
very close to but smaller than the onset of linear stability of the
symmetric soliton. At this saddle-node bifurcation the unstable and
 stable a-symmetric solutions merge to disappear at smaller values
of $\frac{Q}{\sqrt{\kappa }}$.

In the numerical investigations of this problem an interesting
phenomenon comes into play, namely the radiation of energy at large
distances of the solitons. The amount of radiation is stronger when
the initial conditions are further away from a stable solution
\cite{Akh}. This radiation happens in the far wings of the time
dependent amplitude profiles ($\left| u(t)\right|, \left|
v(t)\right|$), where the full equation reduces to its linear part.
Although very strongly dispersive this describes radiation by wave
packets of ever increasing width, but carrying nevertheless energy
and eventually other invariants to infinity. Such a coupling between
a localized system and the infinitely many degrees of freedom of a
radiating field may lead to irreversible effects \cite{pomrotation}.
It shows how subtle may be the distinction between `dissipative' and
`nondissipative' systems as soon as one goes beyond the obvious.
Irreversible process due to radiation may not even require an
infinitely extended physical space. They may also take place in the
reciprocal (or momentum) space by cascade of energy toward smaller
and smaller scales, a typically nonlinear phenomenon
\cite{pomenonlin}. We plan to come back to the issue of the effect
of radiation on quantum phenomena in the present problem. We
shall neglect this kind of effect in the following, since they
cannot be taken into account within our simple formalism. Even
though the radiative losses are present, it was shown by Fadeev and
Korepin \cite {faddeev} that they do not destroy the solitons in a
single fiber, when they are included in the quantized version of the
NLS equation.

In the coming section we shall derive a reduced set of equations
describing the propagation of soliton in coupled fibers. Indeed this
reduction from the original PDE's to a set of coupled ODE's cannot
be quantitatively exact. However with the same choice of trial
functions as Malomed et al.\cite{Malomed} we obtain at least a
reduced system with the right pattern of bifurcation at decreasing
coupling. The fundamental interest of this reduction is that it
allows us to quantize the dynamical system rather straightforwardly.

\section{Classical propagation of solitons in coupled fibers: the
reduced dynamics} \label{sec:classicalproptrial}

Because of the lack of analytical solution in general, we follow an
idea used already by various authors, that allows to understand in a
fairly detailed way the results of the direct numerical simulation
by using an analytical approach. This follows the general method of
research of extrema of functionals by trial functions: dynamics can
be reduced to a minimization problem, then one restricts the
function space where this minimization is done to a space of trial
functions depending explicitly on a few parameters and one studies
the dynamical properties in this reduced space. Since we know the
results of the direct numerical simulations it is in principle
possible to check the quality of the approximation by comparing its
predictions and the `exact' results. This is necessary because the
method of trial functions does not rely on a small or large
parameter and so cannot hope to be `exact' or close to exact in the
usual mathematical meaning of the word. The papers by Malomed et al.
\cite{Malomed} discuss in depth the choice of the trial functions.
We shall not reproduce this discussion here where we take their set
of `optimized' trial functions, following as much as possible their
notations.

The starting point is the writing of the action for the
coupled NLS equations:
\begin{equation}
{\mathcal{S}_{NLS}} = \int_{z_{1}}^{z_{2}} \mathrm{d}z \int
\mathrm{d}t \left[\frac{i}{2}\left(\overline{u}\frac{\partial
u}{\partial z} - u\frac{\partial \overline{u}}{\partial
z}\right)+\frac{i}{2}\left(\overline{v}\frac{\partial v}{\partial z}
- v\frac{\partial \overline{v}}{\partial z}\right)
\\
- \frac{1}{2}|\frac{\partial u}{\partial t}|^2 + \frac{1}{2}|u|^4 -
\frac{1}{2} |\frac{\partial v}{\partial t}|^2 + \frac{1}{2}|v|^4 +
\kappa \left(u{\overline{v}}+ v{\overline{u}}\right)\right]
\mathrm{.} \label{eq:NLSLagrdimcoupl}
\end{equation}

As it can be checked the action ${\mathcal{S}}_{NLS}$ is
proportional to $(z_{2}-z_{1})$ whenever the functions $u(z,t)$ and
$v(z,t)$ are stationary solutions (with respect to the variable $z$)
of the two coupled NLS equations (\ref{eq:coupled1}) and
(\ref{eq:coupled2}) or functions proportional to the same phase
factor $e^{iqz}$. The problem we consider now is how does the
coupling change the propagation of solitons. For that purpose we
reduce the dependence with respect to $t$ to an imposed form (the
trial function) with arbitrary $z$-dependent coefficients, the trial
parameters. Putting this trial form into the action integral and
performing the integration over the variable $t$ yields a functional
of the parameters of the trial function that are themselves
functions of $z$. Doing now the variation with respect to those
functions, one finds at the end a set of ODE's for functions of $z$
only.

The choice of the trial functions is inspired by the soliton
solution in a single fiber and it respects the symmetry between the
two fibers. Following Uzunov et al.\cite{Malomed} one takes:
\begin{equation}
u(z,t) = a(z) \sqrt{\eta(z)} \mathrm{sech}\left[\eta(z) t\right]
\cos(\Theta(z))exp\left[i\left(\Phi(z) + \Psi(z) + q(z)
t^{2}\right)\right] \mathrm{,} \label{eq:trialu}
\end{equation}
and
\begin{equation}
v(z,t) = a(z) \sqrt{\eta(z)} \mathrm{sech}\left[\eta(z) t\right]
\sin(\Theta(z))exp\left[i\left(\Phi(z) - \Psi(z) + q(z)
t^{2}\right)\right] \mathrm{.} \label{eq:trialv}
\end{equation}
In the case of a single fiber carrying a soliton of amplitude $u$,
the trial function $u(z,t)$ becomes the exact one-soliton solution
with $a =\sqrt{\eta }$, $\Theta = q = 0$, $\Phi= z\eta ^{2}/2$ and
$\Psi$ constant. It is important to notice here that the angle
$\Theta$ is for describing the balance between the two fibers,
although the angles $\Phi$ and $\Psi$ have a physical meaning
independent on the trial function, being related to the phase of the
functions $u$ and $v$. The angle $\Theta$ could be replaced by
another parameter in another trial function, not necessarily a
circular function. Inserting this trial form into the action
${\mathcal{S}}_{NLS}$ and performing the integration over $t$, which
is possible because the dependence with respect to $t$ is fully
explicit in the trial functions, one finds a reduced action that is
itself the integral over $z$ of the Lagrange function:
\begin{equation}
{\mathcal{L}} = 2 \kappa a^{2}(z)\cos(2\Psi)\sin(2\Theta) - 2 a^{2}
\cos(2\Theta)\frac{\mathrm{d}\Psi}{\mathrm{d}z}-\frac{1}{3}a^{4}\eta
\sin^{2}(2\Theta) + \frac{2}{3}a^{4}\eta
-\frac{1}{3}a^{2}\eta^{2}-2a^{2}\frac{\mathrm{d}\Phi}{\mathrm{d}z}
-\frac{a^{2}\pi^{2}}{6\eta^{2}}\left(\frac{\mathrm{d}q}{\mathrm{d}z}
+ 2 q^{2}\right) \mathrm{.} \label{eq:trialLagr}
\end{equation}
Up to obvious change in notations (from our $z$ to $\zeta$, from $
\kappa$ to $K$, etc.) this Lagrange function is identical to the one
written by Uzunov et al. \cite{Malomed} but for a misprint in their
paper where the term $q^{2}$ in the last parentheses became $q^{4}$
without harming the rest of their calculation. The parameters of the
trial function are five functions of $z$: $a$, $\Theta$, $\Psi$,
$\eta$ and $q$. The equations of motion for those five functions are
derived by variation of the action, namely the integral over $z$ of
${\mathcal{L}}$. They read:
\begin{equation}
\frac{\mathrm{d}a^{2}}{\mathrm{d}z} = 0 \mathrm{,}
\label{eq:equationa}
\end{equation}
derived by variation with respect to $\Phi$, and

\begin{equation}
\left \{ \begin{array}{l}

\frac{\mathrm{d}\Theta}{\mathrm{d}z} = - 2\kappa \sin(2\Psi)\\
\sin(2\Theta)\frac{\mathrm{d}\Psi}{\mathrm{d}z} =
\frac{a^{2}}{3}\eta \sin(2\Theta)\cos(2\Theta) - \kappa
\cos(2\Psi)\cos(2\Theta)\\
\frac{\mathrm{d}\eta}{\mathrm{d}z} = - 2 q \eta \\
\frac{\mathrm{d}q}{\mathrm{d}z} = - 2 q^{2} +
\frac{2}{\pi^{2}}\left[ \eta^{4} - a^{2}
\eta^{3}\left(1-\frac{1}{2}\sin^{2}(2\Theta)\right)\right]
\label{eq:4eqsmalomed}
\end{array}
\right.
\end{equation}
derived by variation with respect to $\Psi, \Theta, q$, and $\eta$
respectively

\begin{figure}[ht]
\centerline{\includegraphics[scale=0.5]{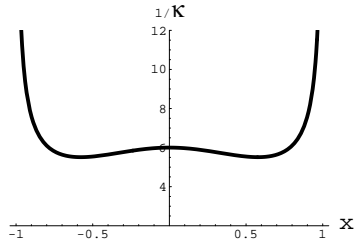}}
 \caption{Inverse of coupling coefficient $\kappa^{-1}$ versus the
stationary values of $x=cos(2\Theta)$. For large coupling, i.e.
$\kappa>1/6$, the symmetric solution, $x=0$, is the only one stable.}
\end{figure}

The parameter $a^{2}$ can be absorbed in the redefinition of
$\kappa$ and will be set to $1$ below (that corresponds to a mass
$Q=2$).

 The soliton solutions are $z$-independent solutions of this set of equations. There are two classes of soliton solutions in this model,
depending on the coupling parameter. For any coupling there exists a
symmetric soliton, with equal intensity in both fibers, i.e. $\Theta
= \frac{\pi}{4}$. At small coupling this symmetric solution is
unstable against asymmetric soliton. Such an asymmetric soliton is
found by canceling the  $z$-derivatives in equations
(\ref{eq:4eqsmalomed}) and  choosing $\cos(2\Psi) = 1$. This yields
the relation between the coupling coefficient $\kappa$ and the
 balance parameter for the intensity $x=\cos2\Theta$
\begin{equation}
\kappa^{-1}=\frac{6}{(1+x^{2}) \sqrt{1-x^{2}}}
\mathrm{,}
\label{eq:subcritic}
\end{equation}

which is illustrated in
  Fig.1. Using the trial functions, the bifurcation in the set of possible solutions
  is found to occur at
the critical value, $\kappa_{c}=1/6$ which is 11 per cent less than
the exact value \cite{Akh}, $\kappa_{c}=\sqrt{3}/4$ . Moreover the
bifurcation is slightly subcritical, in good agreement with the NLS
results \cite{Malomed}. Close to the bifurcation point there is a
small range of values of the coupling, $\frac{1}{6}\leq \kappa \leq
\frac{4}{9\sqrt{6}}$ ($0.167\leq \kappa \leq 0.181$) where there are
three sets of solutions: the symmetric solution that is linearly
stable, and two pairs of asymmetric solutions, one linearly stable
and another linearly unstable. The branch of stable asymmetric
solution does not merge smoothly with the symmetric solution, but
disappear when it has still a finite amplitude. The main conclusion
that we shall draw here is that this set of trial functions
reproduces well the pattern of bifurcation of the exact model. This
makes it a good candidate for studying the quantum tunneling.

Before to start this study, let us explain how we managed to define
 a quantity related to the usual potential energy of a
mechanical system. Although this is not strictly necessary it helps
to draw various quantities relevant for analyzing the tunneling by
making a connection, however loose it is, with the familiar notions
of barrier and of barrier crossing.

The `potential energy' is derived from the total energy associated
to the dynamical system under consideration, namely the equations
(\ref{eq:equationa}) to (\ref{eq:4eqsmalomed}). An expression for
this energy is given by Uzunov et al. With $a^{2} =1$ it becomes:
\begin{equation}
{\mathcal{H}}_{\mathrm{trial}}^{\mathrm{cl}} = - 2 \kappa
\cos(2\Psi)\sin(2\Theta) + \frac{1}{3} \eta \sin^{2}(2\Theta) -
\frac{2\eta}{3} + \frac{\eta^{2}}{3} +
\frac{\pi^{2}q^{2}}{3\eta^{2}} \mathrm{.} \label{eq:energtrial}
\end{equation}
Note that the Lagrangian (\ref{eq:trialLagr}) includes terms linear
with respect to  first derivatives (with respect to $z$). It means
that the two successive  operations of choosing trial functions and
averaging over the retarded time $t$, lead from the Lagrangian
formalism to the Hamiltonian one, with
\begin{equation}
{\mathcal{S}}= \int (\sum_{i=1,2} p_{i}dq_{i}- {\mathcal{H}}dz)
\label{eq:lag}
\end{equation}
The first term in the r.h.s. of equation (\ref{eq:lag}) will be the
one responsible for the Euclidian action derived in the next
section. As already noticed by Uzunov et al., equations
(\ref{eq:equationa}) to (\ref{eq:4eqsmalomed}) are the Hamilton
equations of a two-degrees of freedom system. The two pairs of
conjugate variables are $\{\Psi ,2x=2\cos (2\Theta )\}$ , and
$\{q,y=\frac{\pi ^{2}}{6\eta ^{2}}\}$ , i.e. the phase and amplitude
differences, as well as the chirp and width, respectively. We are
interested in the value of ${\mathcal{H}}$ for steady states, that
turns out to be a simple function of the coupling $\kappa$. That
should give an idea of how the energy changes when the variables are
different of their values in the steady state(s). In order to
preserve the connection with a potential energy in the usual sense
we impose that, at the equilibrium points, this `potential' energy
is at an extremum. This is realized (probably not uniquely) by
plugging into ${\mathcal{H}}$ the values of $q$ and $\Psi$ at the
various equilibria, that is $\Psi = 0$ and $q = 0$ to cancel the
conjugate momenta. This yields:
\begin{equation}
{\mathcal{H}_{pot}} = - 2 \kappa \sin(2\Theta) + \frac{1}{3} \eta
\sin^{2}(2\Theta) - \frac{2\eta}{3} + \frac{\eta^{2}}{3} \mathrm{.}
\label{eq:energtrialpotent}
\end{equation}

\begin{figure}[ht]
\begin{center}
(a)\includegraphics[scale=0.5]{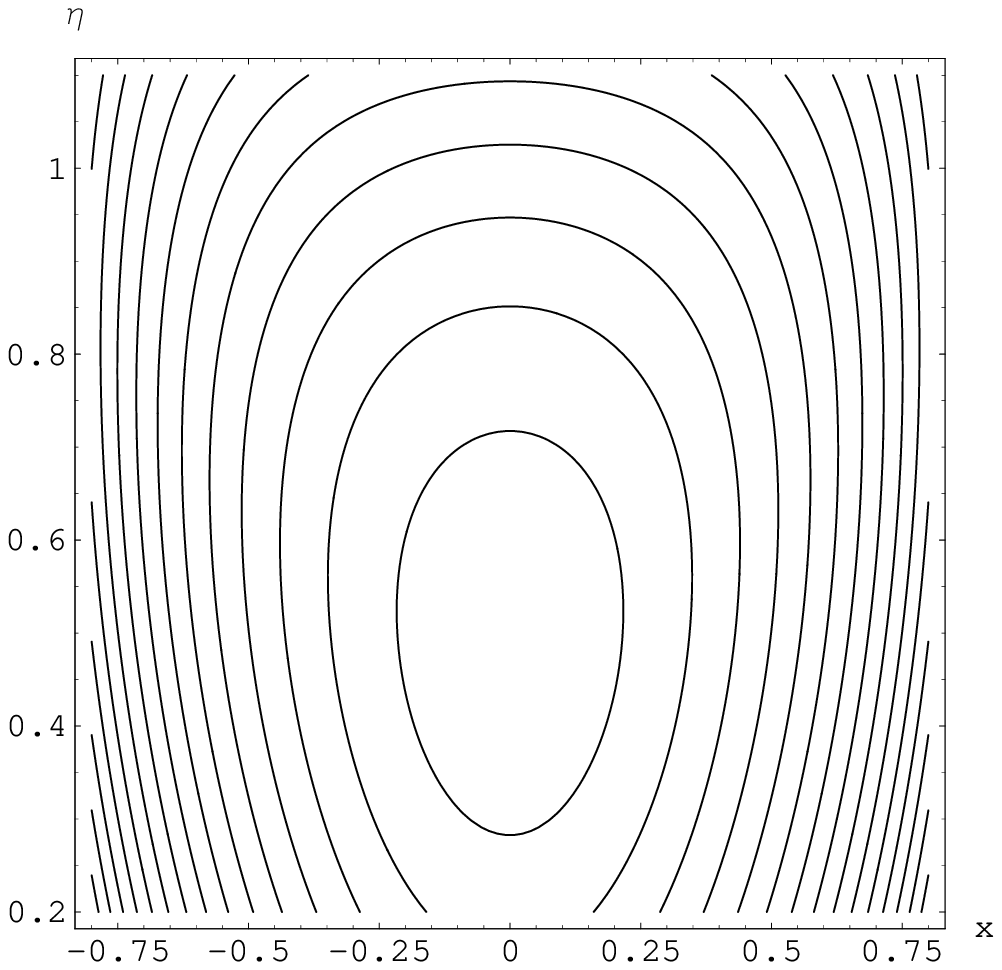}
(b)\includegraphics[scale=0.5]{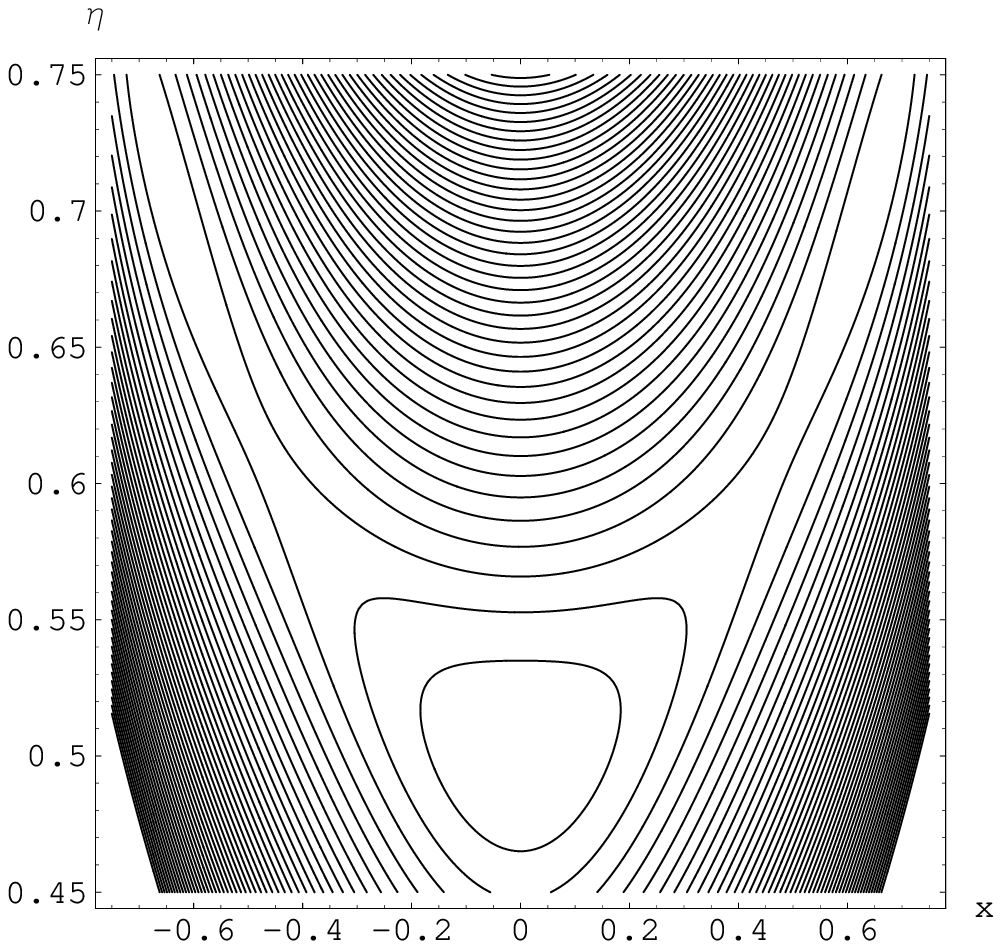}
 (c)\includegraphics[scale=0.43]{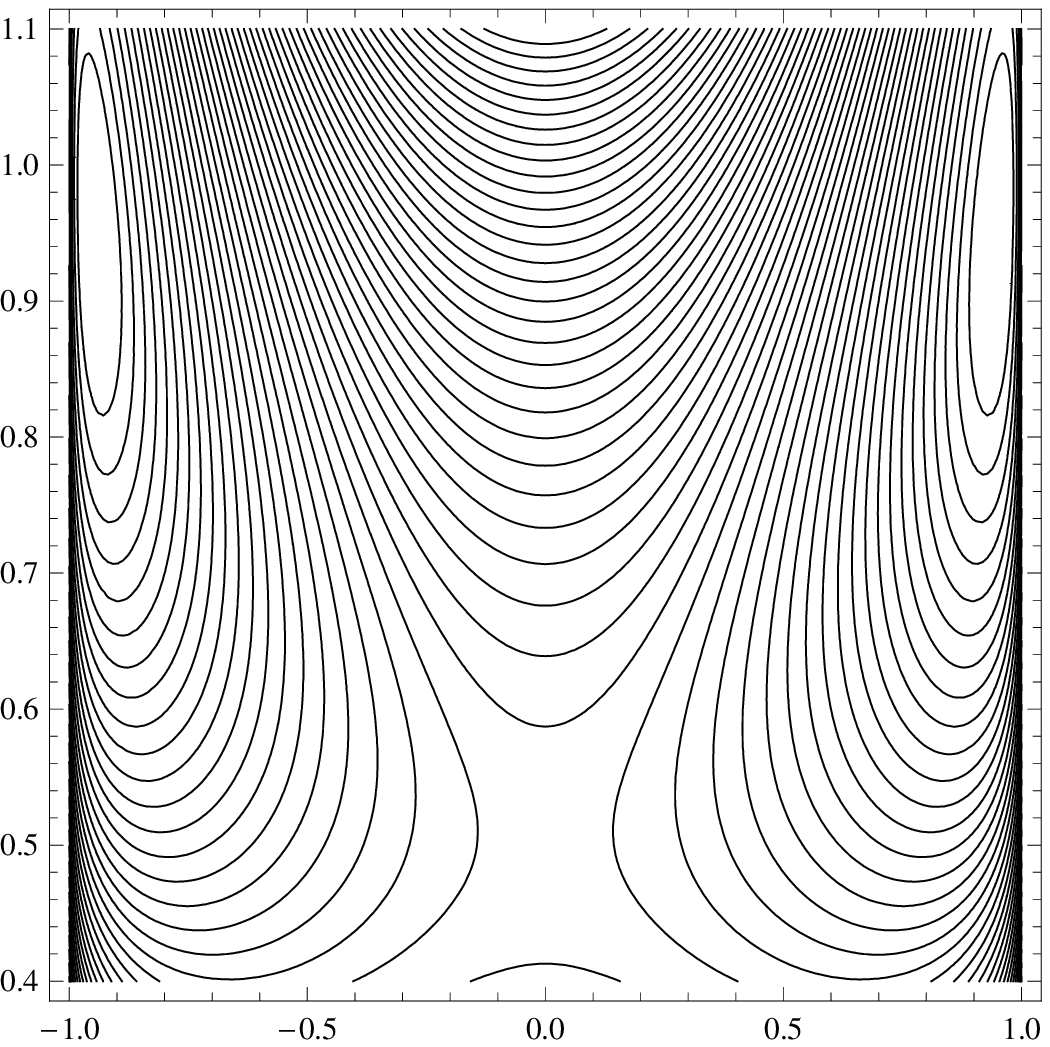}
\end{center}
 \caption{Level lines of the potential in the plane ($x,\eta$), (a) for
 $\kappa=0.5$, only the symmetric solution $x=0$ is stable, (b)for
 $\kappa= 0.18$ which belongs to the subcritical domain in Fig.1 where symmetric and
a-symmetric solutions are stable, (c) for $\kappa=0.1$, only two
asymmetric solutions are stable}
\end{figure}

This `potential' energy depends on two parameters, $ x=\cos (2\Theta
)$ and $\eta $, and it is plotted in Fig. $2$ for various coupling
strength to show the bifurcation of the equilibria from a single
equilibrium at large coupling, Fig. $2(a)$,  to a more complex
pattern, as the coupling decreases. In particular, some sort of
barrier is evident in Fig. $2(c)$. It separates the two deep minima
of the potential lying each in the vicinity of  $(\pm 1,1)$. Each
minimum corresponds to one of the stable asymmetric soliton,
although the unstable symmetric soliton at $(0,1/2)$ is a saddle
point of the potential energy.

The above picture illustrates the known results: classically there
is no way for a soliton initially in a given fiber, to escape
through the other fiber, at low coupling, because of the barrier.
Before to present the quantum version of this problem, let us precise
what is the low coupling range in terms of physical quantities.
Note first that the low-coupling range writes
\begin{equation}
\frac{a ^{2}}{\kappa} > 6
 \mathrm{,}
 \label{eq:seuil}
\end{equation}
 for an incident soliton of the form $u(0,t)=a/ch(at)$ injected in one of the two fibers
 ($a=1$ above). Secondly let us define the scaling quantities in
equations(\ref{eq:coupled1})-(\ref{eq:coupled2}), by
 using the soliton units, $z=z^{phys}/L_{D}$, $t=t^{phys}/\tau_{0}$, $u,v=\sqrt{\frac{2\pi n'_{2}}{\lambda_{0}} L_{D}}B_{1,2}$, and
$\kappa=\kappa^{phys}L_{D}$ ,  where
$L_{D}=\frac{\tau_{0}^{2}}{k"_{0}}$ is the dispersion length,
 and $B_{1,2}$ are the
slowly varying amplitudes of the electric field (see appendix). The
relation (\ref{eq:seuil}) becomes

\begin{equation}
n_{2}I_{M}= n'_{2}\left| {B_{M}^{2}}\right| > \frac{3}{\pi}
\lambda_{0}\kappa^{phys}
 \mathrm{,}
 \label{eq:seuilphys}
\end{equation}

or, $n_{2}I_{M} > \frac{3}{2}\frac{\lambda_{0}}{L_{c}}$ when
introducing the switching length $L_{c}=\frac{\pi}{2\kappa^{phys}}$
defined for the CW linear regime. Using the relation
(\ref{eq:beta}), the low coupling range also writes

\begin{equation}
L_{c}> 3 \pi L_{D} \label{eq:Lcseuil} \mathrm{.}
\end{equation}

\section{Semiclassical quantization of the coupled fiber system}
\label{sec:quantization}

Before computing the quantum tunneling, let us recall the main
differences between the classical solitonic solution and its
quantized form. In quantum mechanics a state localized on one side
or on the other only is not an eigenstate of the system, because of
the possibility of tunneling. Therefore if one starts at `time' zero
with all the amplitude on one side (meaning all the probability in
one of the two possible asymmetric states), after the time of
tunneling this will be transferred to the other side and eventually
oscillate between the two sides. It is also possible to inject at
the input of the dual core fiber, the quantum ground state, which is
symmetrical, with equal amplitude in the two sides.  To have a
physical image of the process by which the transition occurs between
the two asymmetric states, one may recall that the number of photons
is not fixed in the quantized soliton, so that it fluctuates in both
fibers. Therefore, the fluctuations may bring one fiber into the
soliton state, although the other goes to the state without soliton,
and the two states switch in the course of time, as studied below.

 Let us now outline how to compute the quantum tunneling
between the two fibers. Because we have a classical field, the
quantization of the coupled equations (\ref{eq:coupled1}),
(\ref{eq:coupled2}) for the two fibers belongs to the general
problem of quantization of field theories. Although this may be done
formally, it requires a rather heavy machinery in any case.
Fortunately there are various possible short cuts in this
derivation. The most obvious one is to reduce PDE's system to a set
of ODE's, by using trial functions depending on a certain set of
unknown parameters. By refining the choice of trial functions ad
infinitum, namely by introducing trial function with more and more
parameters, one should converge in principle toward the exact
result. But we will merely use the above described trial functions.
The Euler--Lagrange condition of stationarity of the action yields a
set of dynamical (in `time' $z$) equation, that can be formally
quantized because it has a symplectic structure.

This is what we are going to do, except for one point. It is
possible to short cut all this explicit quantization in the WKB
limit, where the wave function is expressed by means of the
classical Hamilton-Jacobi action, $\Phi= A\exp(i\mathcal{S}/\hbar)$.
This is the well-known quasi--classical limit, that restricts
oneself to situations where any action involved is typically much
bigger than $\hbar$ . This seems a reasonable limit, but it does not
necessarily cover all possible situations-we shall come at the end
to what seems to be `the' standard experimental situation in this
respect. The WKB limit is especially convenient for treating
tunneling problems, because it amounts to calculate the imaginary
part of the action (which is complex) and to put at the end $\hbar$
at the right place. Indeed the tunneling factor is is given by $T =
exp(-2 S_{E}/\hbar)$, at leading order. Here $S_{E}$ is the
imaginary part of the action, which enters then in the modulus of
the wave function as a real exponent (instead of the usual imaginary
exponent relevant for the classical limit of quantum mechanics).
This imaginary part of the action is calculated by two steps. First
one has to change the conjugate variables $(q,p)$ into $(q, ip)$ in
the classical Hamilton-Jacobi formulation of quantum mechanics, the
Hamiltonian $H(q, p)$ becoming $H(q, ip)$. Secondly one is left with
a problem of extremalization of a new action, the Euclidean action,
that is formally another problem of classical mechanics. For
instance in the often presented problem of a particle of energy
$\emph{E}$ in a double well potential $V(q)$, with Hamiltonian
$H=\frac{p^{2}}{2m} + V(q)$, the Euclidean action is calculated with
the abbreviated action \cite {landau}

\begin{equation}
\mathcal{S}_{E}=\int_{q[0]}^{q[z_{f}]}pdq \label{eq:action-red}
\end{equation}.

derived from the Hamiltonian $H_{E}= -\frac{p^{2}}{2m} + V(q)$  but
with the same energy as the one of the classical motion. For the
potential this means that it gets rotated by 180 degrees, thus
exhibiting two "hills" of maximal energy. The values of $q[0]$ and
$q[z_{f}]$ in equation (\ref{eq:action-red})  are those of the
classical turning points defined by $\emph{E}=V(q)$ . To calculate
$\mathcal{S}_{E}$, one has to find a trajectory joining these
points, namely to calculate an Euclidean path integral. This is
performed by solving the Hamilton equations for the Euclidean
Hamiltonian

\begin{equation}
\left \{ \begin{array}{l}

\frac{\partial q}{\partial z} =\frac{\partial H_{E}}{\partial p}\\
\frac{\partial p}{\partial z} = - \frac{\partial H_{E}}{\partial q}\\
\label{eq:pq}
\end{array}
\right.
\end{equation}

 by taking as initial conditions, the known value $q[0]$, and an unknown value $p[0]$.
  By varying the latter value, one finally converges towards a trajectory ending
at $q[z_{f}]$,
 which provides the action defined in equation
 (\ref{eq:action-red}).
 Note that equations (\ref{eq:pq}) are obtained from the
classical Hamiltonian system ( which is identical to equation
(\ref{eq:pq}) but with $H$ in place of $H_{E}$), by changing $z$ in
$iz$, and $p$ in $ip$. The change to an imaginary "time" ( from $z$
to $iz$ here) amounts to go from a Minkowskian to an Euclidean
metric. Therefore equations (\ref{eq:pq}) are called "Euclidean
equations of motion", and their classical solution joining the two
"vacua" of the double-well potential, often named "kink solution",
is an example of an instanton \cite{coleman} in quantum mechanics.

In the above example the variables (p, q) are the impulsion and
position of a particle in a 1D potential. Generalization to cases of
a multidimensional set of generalized coordinates and momenta leads
to similar relations \cite{landau}.

\subsection{Semi-classical Action.}
\label{subsec:semiclassaction}

To put all those principles in practice we have to formalize the
dynamical system (equation (\ref{eq:4eqsmalomed})) in terms of
canonically conjugate variables. Once this is done, the Euclidean
equations of motion are found by multiplying the "time" $z$ and the
momenta by $i$. As noted in section III, the reduced equations
(\ref{eq:4eqsmalomed}) are those of an Hamiltonian system with two
degrees of freedom, therefore a simple choice for conjugate
variables $(q_{j},p_{j})$ with $j=1,2$, is to take the pair
$(x=cos(2\Theta)\mathrm{,} y=\frac{\pi ^{2}}{6 \eta ^{2}})$ as
coordinates and $(2\Psi\mathrm{,} q)$ as their conjugate momenta.
The Euclidean Hamiltonian is obtained from the classical one in
equation (\ref{eq:energtrialpotent}), by changing $\cos(2\Psi)$ into
$\cosh(2\Psi)$, and $q^{2}$ into $-q^{2}$. It becomes

\begin{equation}
{\mathcal{H}}_{\mathrm{E,trial}} = - 2 \kappa
\cosh(2\Psi)\sin(2\Theta) + \frac{1}{3} \eta \sin^{2}(2\Theta) -
\frac{2\eta}{3} + \frac{\eta ^{2}}{3} - \frac{\pi^{2}
q^{2}}{3\eta^{2}} \mathrm{.} \label{eq:quenergtrial}
\end{equation}

The semi-classical dynamics is then driven by the new set of four
(Euclidean) equations, that are the Hamilton equations for the
conjugate variables $(q_{j},p_{j})$, deduced from the Euclidean
Hamiltonian (\ref{eq:quenergtrial})

\begin{equation}
\left \{ \begin{array}{l}

\frac{\mathrm{d}x}{\mathrm{d}z} =-2\kappa \sinh (2\Psi)\sqrt{1-x^{2}}\\
\frac{\mathrm{d}(2\Psi)}{\mathrm{d}z} =\frac{2}{3}\eta x-2\kappa  \cosh (2\Psi)\frac{x}{\sqrt{1-x^{2}}}\\
\frac{\mathrm{d}\eta}{\mathrm{d}z} =2q\eta\\
\frac{\mathrm{d}q}{\mathrm{d}z} =2q^{2}+\frac{2}{\pi ^{2}}(\eta^{4}-
\eta ^{3}\frac{1+x^{2}}{2}) \label{eq:4eqs}
\end{array}
\right.
\end{equation}

\begin{figure}[h]
\centerline{\includegraphics[scale=0.5]{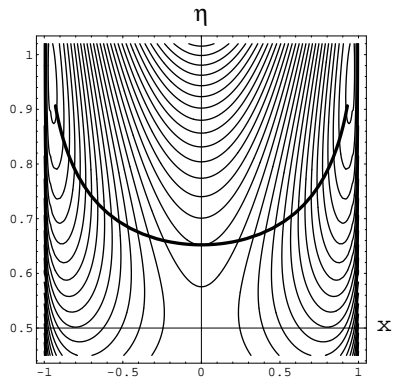}} \caption{ Quantum
trajectory superposed to the potential for $\kappa=0.1$: only two asymmetric solutions are stable.}
\end{figure}

As in the case of a particle in a double well potential, calculating
the  probability for the soliton to tunnel through a classically
forbidden region ( $H_{pot}$ ) with the Minskowskian space path
integral, corresponds to calculating the transition probability to
tunnel through a classically allowed region ( $-H_{pot}$ ) in the
Euclidean path integral, with the action

\begin{equation}
\mathcal{S}_{E}=\int_{q_{L}}^{q_{R}}(p_{1}dq_{1}+p_{2}dq_{2})
\label{eq:action2}
\end{equation}.

where $q_{L},q_{R}$ are the coordinates of the turning points. To
perform the integration giving the action, it is enough to choose a
convenient integration path in the Euclidean plane connecting the
two minima $M(x_{M},\eta_{M})$ of the classical potential in Fig.
$3$, which become the maxima of the Euclidean potential. For small
values of $\kappa$, one has ${\mathcal{H}}_{\mathrm{pot,M}} \approx
-\frac{1}{3}(1+18\kappa ^{2})$. In the present case it is easier to
carry the integral from $x_{0}=0$ up to $x_{M}$. We set the value of
the Hamiltonian $\mathcal{H}_{0}$ close to
${\mathcal{H}}_{\mathrm{pot,M}}$. Because of the symmetry of the
heteroclinic trajectory joining the two extrema of the potential, we
have to choose the initial condition $q_{0}=0$. Then the initial
value of the phase difference $\Psi_{0}$ is deduced from equation
(\ref{eq:quenergtrial}), and we have only one initial parameter to
adjust, $\eta_{0}$, in order that the trajectory ends with a
vanishing impulse $q_{f}=\Psi_{f}=0$ close to the extrema
$M(x_{M},\eta_{M})$ in the plane $(x,\eta)$. The integration path is
shown in Fig.$3$. The action along the semi-classical trajectory is
given by the expression (\ref{eq:action2}) that writes with our
notations
\begin{equation}
{\mathcal{S}}_{E}= 2\left|\int_{z=0}^{z=z_f} \left[-4\Psi(z)\kappa
\sinh
(2\Psi(z))\sqrt{1-x(z)^{2}}-\frac{2\pi^{2}}{3}\frac{q^2(z)}{\eta^2(z)}
\right]\mathrm{d}z\right| \mathrm{.} \label{eq:actionE}
\end{equation}

\begin{figure}[h]
\centerline{\includegraphics[scale=0.7]{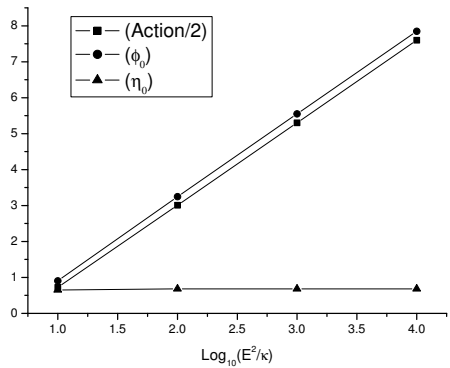}} \caption{
Action/2 (squares), Maximum of the impulse phase $\Phi_{0}$
(circles), and $\eta_{0}$ (triangles) versus the scaled coupling
parameter }
\end{figure}

The numerical result of the integration is shown in Fig. $4$ which
displays the action as a function of the coupling parameter $\kappa$
in a logarithmic scale.  In the domain of existence of the
asymmetric  solution, $\kappa^{-1}>6$ in Fig. $1$, the action
clearly displays a logarithmic dependence with respect to the
coupling, we have the law

\begin{equation}
\mathcal{S}_{E}=2\ln (\kappa _{c} /\kappa) \label{eq:lnk}
\end{equation}
that holds true with a precision better than $1$ per cent over many
decades, with the numerical value $\kappa_{c}=0.2$ slightly higher
than the bifurcation one $0.18$. Note that the relation
(\ref{eq:lnk}) holds true except in the close  vicinity of the
bifurcation point, not visible in Fig. $4$. The $ln$-dependence in
equation (\ref{eq:lnk}) follows   straightforwardly from the
substitution of exponentials for the hyperbolic sine in the equation
of motion for Euclidean dynamics. We also report in Fig. $4$ the
dependence of $\Psi_{0}$ and $\eta_{0}$ as function of $\kappa$. At
$x=0$  the solution becomes transiently symmetric,
$\sin(2\Theta)=0$, but its width is different from the symmetric
value, $\eta_{0}\neq \frac{1}{2}$, and the impulse $\Psi_{0}$ is
maximum. We show that $\Psi_{0}$ evolves very much as the action,
while $\eta_{0}$ is quite constant. Actually the heteroclinic
trajectory drawn in Fig.(3) passes through the abscissa $x=0$
approximately at the ordinate $\eta\sim 0.67$ whatever the value of
the coupling constant, while the impulse here increases like $\ln
(\kappa_{c} /\kappa)$.  This result shows the  leading role of the
conjugate variables $\Psi$ and $x=\cos(2\Theta)$ in the dynamics. At
this stage it is interesting to compare the latter result
(\ref{eq:lnk}) with the action derived by a simpler choice of trial
functions, based on the hypothesis of constant width soliton (and of
no chirp), as proposed by Par\'e \cite{Pare} and Kivshar
\cite{Kivshar}. In these simpler cases, one obtains a single degree
of freedom Hamiltonian dynamics. The approximate calculation of the
Euclidean action may be done  analytically, and leads to similar
results in both cases. With the notations of Kivshar, for example,
using as conjugate variables $(\Phi, \Delta)$, the calculation of
the action amounts to carry the integral
$\mathcal{S}_{E}=\int_{-1}^{+1}\Phi (\Delta )\mathrm{d}\Delta $, the
function $\Phi (\Delta )$ being given explicitly in \cite{Kivshar}.
In the limit of a small
   coupling and with $\kappa=\gamma^{-1}$, the equation for $\Phi$   reduces, at leading order,
   to $\Phi \approx i\ln(\gamma) + \hat{\Phi}$ where $\hat\Phi$ is
   the solution of $\frac{1}{I(\Delta)} = e^{i\hat{\Phi}}$ that is
   of order $1$. Therefore in this limit $\gamma$ large (equivalent to
   small coupling), $\Phi \approx
   i\ln(\gamma)$ so that the action associated to tunneling is just
   $S \approx 2 i\ln(\frac{\gamma}{\gamma_{c}})$, where $\gamma_{c}  $ is a constant.

Summarizing the Euclidean action obeys the law (\ref{eq:lnk}) in all
cases of trial functions we have considered, i.e. for a single
degree of freedom Hamiltonian as well as with two degrees of
freedom. Consequently, it does not seem necessary to refine more
our model to obtain the information we need, i.e. the order of
magnitude of the tunneling amplitude.

\subsection{Tunneling factor.}
The possibility for the soliton to tunnel from one fiber to the
other in real space, is measured by the transmission coefficient, with the expression
  $T = \left| \frac{F}{A}\right| ^{2} $ in
a double well tunneling problem, with $F,A$ the amplitudes of the
transmitted and  incident waves, respectively \cite{griffiths} . It
has already been noted that  the transmission is given by $T=exp(-2
S/\hbar)$ at leading order. In practice $S$ is the "physical
action", having the same dimension as $\hbar$. Therefore, to
calculate the "true" transmission for the soliton in the two coupled
fibers one has to multiply the dimensionless action $S_{E}$ by an
appropriate coefficient $s^{(1)}$ depending on the properties of the
fiber and of the characteristics of the EM wave, this giving lastly
the "physical action" $S_{E}^{phys}=s^{(1)}S_{E}$ which has the
dimension of $\hbar$. As shown in the appendix

\begin{equation}
2 s^{(1)}/\hbar=\gamma/(\omega_{0}\tau_{0})^{3} \label{eq:S1}
\end{equation}

where
\begin{equation}
\gamma\sim 8\sigma\varepsilon _{0}c^{3}k"_{0}^{2}/(n'_{2}\hbar)
\mathrm{.} \label{eq:gam}
\end{equation}

The value of $\gamma$ depends on the fiber parameters $n'_{2}$ and
$\sigma$, cross section of the fiber. Let us consider  $10\mu m^{2}$
area silica fibers with  $n'_{2}=2.6 10^{-22}(m/V)^{2}$ as given in
(\cite{erlangen}), (\cite{newell}). With the values of coefficients
given in the appendix in MKS units, the coefficient $\gamma$ is
about $3.6 10^{8}$ .

With equations (\ref{eq:lnk})-(\ref{eq:gam}), the transmission  coefficient

\begin{equation}
T=exp-[\frac{2\gamma}{(\omega_{0}\tau_{0})^{3}}
ln(\kappa_{c}/\kappa)]
\mathrm{,}
\label{eq:tunnel}
\end{equation}

or
\begin{equation}
T=(\frac{\kappa}{\kappa_{c}})^{\frac{2\gamma}{(\omega_{0}\tau_{0})^{3}}}
\mathrm{,}
\label{eq:tun}
\end{equation}

behaves as a power law, that is smoother than the usual
exponential in tunneling amplitudes. The tunneling is possible when the exponent in
equation (\ref{eq:tunnel}) is "not too big".  In the semi-classical
regime considered above, the phase of the wave-function is derived
by expansion at lowest order with respect to $\hbar$. This requires
that the exponent $\ln(T)= \frac{2\gamma}{(\omega_{0}\tau_{0})^{3}}
ln(\kappa_{c}/\kappa)$ is much larger than unity, then the
probability of tunneling is obviously weak. When the exponent
becomes smaller or of order unity, one is in the "pure quantum
limit", and the previous derivation is no more valid, since the
wave-function cannot reduce to its first order term in $\hbar$.
Nevertheless we can assert by continuity argument that tunneling
continue to exist, and that it is likely much more efficient. The  boundary
between these two limits can be defined by

\begin{equation}
\frac{\gamma}{(\omega_{0}\tau_{0})^{3}} ln(\kappa_{c}/\kappa)=1
\mathrm{.}
\label{eq:frontiere}
\end{equation}

This dependence is drawn in Fig. $5$ for the value of $\gamma$ given
above. The quantum regime is reached as soon as the pulse duration
is longer than a  $ ps$. Therefore quantum tunneling seems within
reach of present days experiments.

\begin{figure}[h]
\centerline{\includegraphics[scale=0.6]{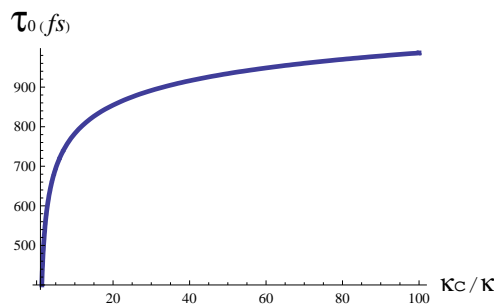}}
\caption{Boundary between the quantum and semi-classical regime,
 for a silica core fiber. The
quantum regime  stays above the frontier.}
\end{figure}

\bigskip
\subsection{Quantum switching.}
To estimate the typical length needed for the soliton to
 tunnel from fiber to fiber, we reason as
follows. We estimate first the time scale for the quantum tunneling. We split the
wavefunction into the `right' amplitude, $\Phi_{R}$, and the
left one, $\Phi_{L}$, each one being for the state in
one fiber only. Because of the tunneling those states are not
eigenstates but split into two eigenstates, one even (the ground  state)
$\Phi_{S}$ and the other odd, $\Phi_{A}$, under permutation of the
two fibers. One has

$\Phi_{L}=(\Phi_{S}-\Phi_{A})/\sqrt{2}$, and
$\Phi_{R}=(\Phi_{S}+\Phi_{A})/\sqrt{2}$.

The energy difference between the symmetric and antisymmetric state
gives, via the Planck-Einstein relation, the typical tunneling time.
Let $A$ be half of this energy difference. If at time zero the soliton is on the right fiber, the evolution of
its amplitude later on is given by

\begin{equation}
\Phi(t)=\frac{1}{\sqrt{2}}(\Phi _{S}e^{-iE_{S}t/\hbar }+\Phi
_{A}e^{-iE_{A}t/\hbar }) \mathrm{.}
\end{equation}

Therefore the amplitude in one fiber oscillates with the period

\begin{equation}
T_{osc}\sim h/2A
\mathrm{.}
 \label{eq:period}
\end{equation}

In the following derivation, we approximate the energy splitting in
each well by using the standard result for a particle of
momentum $p(x)$ in a double-well:

\begin{equation}\label{A}
2A=\frac{\hbar \omega}{\pi} exp(-\frac{1}{\hbar }\int_{-a}^{a}\left|
p\right| dx)
\end{equation}

where $[-a, a] $ is the $x$-range under the barrier for the given
energy, and $\omega$ the pulsation of the wave-function in the
bottom of the well. For a quadratic potential $V(x)$,
of curvature $V"$ around the minimum $x_{M}$,
the pulsation of a particule of mass $m$ is such that

\begin{equation}\label{eq:courb}
V"= m \omega^{2}
\end{equation}
The mass of the particle is deduced from its momentum under the
barrier of height $U_{0}$, at $x=0$, where

$ p_{0}^{2}/2m = U_{0}$. Therefore the pulsation writes

\begin{equation}\label{eq:omeg}
\omega=\sqrt{ \frac{2 V" U_{0}}{p_{0}^{2}}}
\end{equation}

where all quantities are in physical units. Note that the dimensions
are $[P][x]=[S]$ and $[V"]=[V]/[x^{2}]$, therefore the dimension of
$x$ plays no role. For the fiber problem, we shall  consider only
one set of conjugate variables, ($x,\Psi$),
neglecting the $\eta$ dependance of the potential, which plays a   secondary
role, moreover the physical quantities in equation (\ref{eq:omeg})  have
to be expressed in terms of the reduced ones, and the "time" period
of equation (\ref{eq:period}) becomes a spatial period. This writes

$$\left \{ \begin{array}{l}

p_{0}=S^{(1)}\Psi_{0}\\
V"=W^{(1)}v" = \omega_{0}S^{(1)}v"\\
U_{0}=W^{(1)}\Delta V\\
T_{osc}= Z k_{0}^{"}/\tau_{0} \\
\mathrm{,}
\label{eq:corr}
\end{array}
\right.$$

where the curvature $V"$ and the height $\Delta V$ of the potential
barrier are deduced from equation (\ref{eq:energtrialpotent}), that
gives

  $\Delta V =
0.037+6\kappa^{2}$, and $v"= \frac{2}{3^{3}}\kappa^{-2}$. Moreover
the numerical results in Fig. $4$ give
$\Psi_{0}=ln(\kappa_{c}/\kappa)$. In the above relations $Z$ is the
"true" spatial period along the optical axis of the fibers, obtained
from equation (\ref{eq:NLS}) after dividing all terms by
$k_{0}^{"}/\tau_{0}$ to obtain a soliton of half-mass equal to unity
as assumed in the present section.

With these expressions,  the  equation (\ref{eq:period}) becomes

\begin{equation}\label{omeg}
\omega_{0} T_{osc}= Z \frac{\omega_{0} k_{0}^{"}}{\tau_{0}}=\frac{2
\pi^{2}\Psi_{0}} {\sqrt{2v"\Delta V}} \exp{S/\hbar}
 \mathrm{,}
\end{equation}

with $S/\hbar=\frac{\gamma}{(\omega_{0}\tau_{0})^{3}}
\ln(\kappa_{c}/ \kappa)$.

\begin{figure}[ht]
\begin{center}
(a) \includegraphics[scale=0.6]{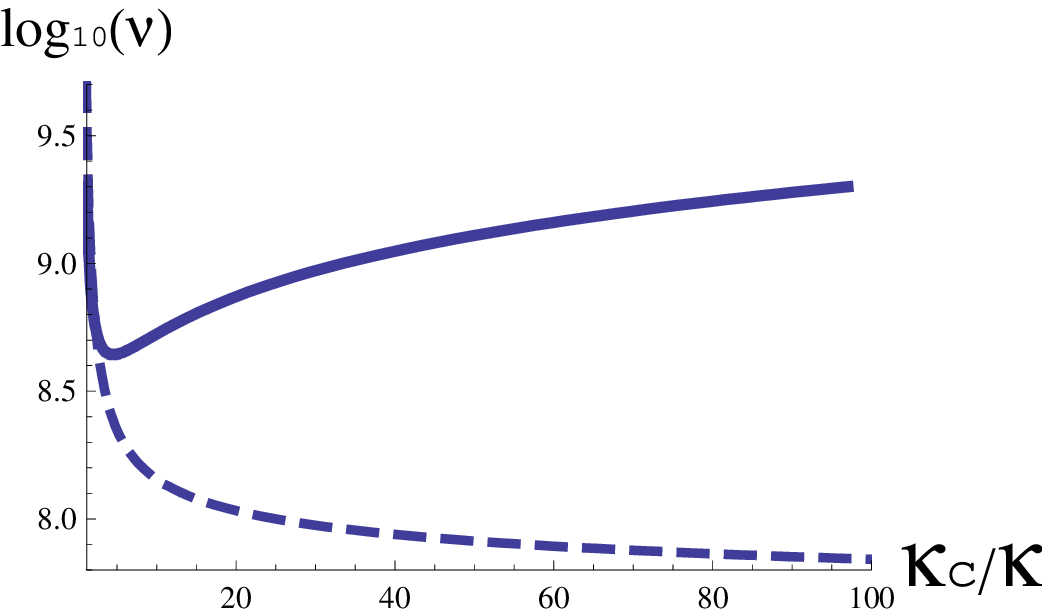}
(b)\includegraphics[scale=0.6]{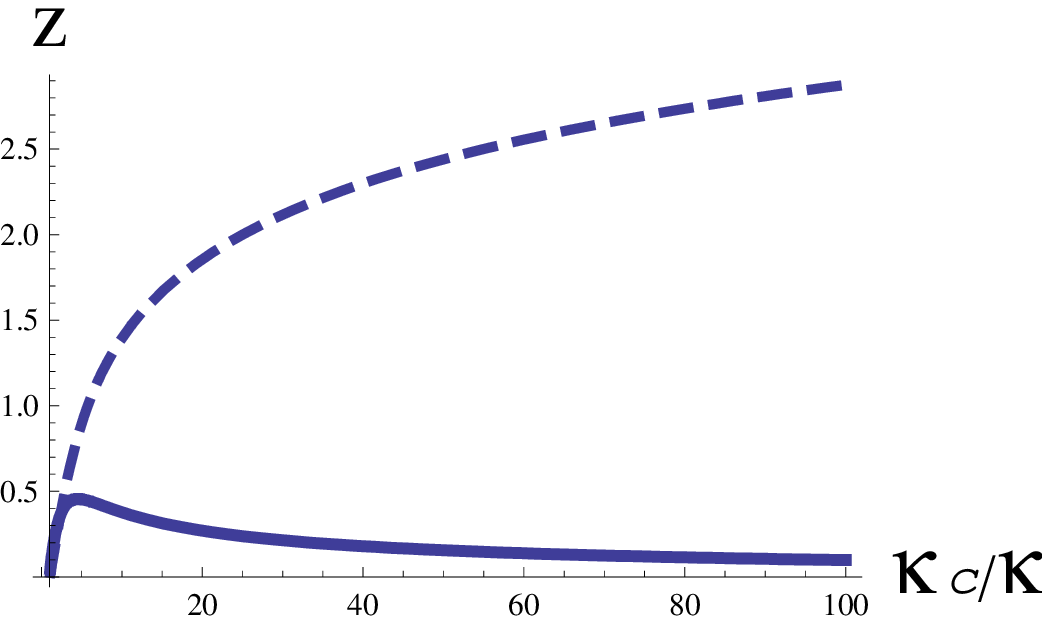}
 \end{center}
 \caption{(a) Frequency of the periodic switching $\nu=1/\tau$, in Log scale, with $\nu$ in $Hz$;
 (b) Spatial period $Z$, in $m$,  as function of the ratio $\kappa_{c}/\kappa$, for the case
 $\tau_{0}=0.6
\emph{ps}$ (r=1, dashed line) and $1.3 \emph{ps}$ (r=0.1, solid
line). For
 $\frac{\kappa_{c}}{\kappa}=100$, the period is $Z=10cm$ for the solid line, and about $3m$ for the dashed line .}  \label{fig:fignuZ}
\end{figure}

Since the probability of finding the soliton in a given fiber
oscillates with respect to the spatial variable $z$ with a
wavelength  $Z$, it also oscillates in time from one fiber to the
other with the period $\tau = nZ/c$, at a given $z$. Using the
numerical values given in the appendix for standard fibers, the
period of the switching depends on the two parameters $\kappa$ and
$\tau_{0}$. The frequency $\nu=c/nZ$, and the spatial period $Z$ are
drawn in fig.(\ref{fig:fignuZ}), as function of the coupling
parameter ratio $\kappa_{c}/\kappa$. We have chosen two values of
pulse duration, $\tau_{0}= 0.6 ps$ (dashed line, corresponding to
$r=1$), and $1.3 ps$ (solid line, $r=0.1$), which are respectively
below, and above the frontier drawn in Fig.(5). More precisely the
dashed line stands into the "semiclassical" regime as soon as
$\kappa_{c}/\kappa$ is larger than few units where the WKB
approximation is valid, whereas the solid line corresponds to the
"purely quantum" regime. The two lines displays high frequencies,
ranging from hundred of $Mhz$, towards tens of $Ghz$, that could be
interesting for applications to high speed transmission. Note that
while the solid line corresponds to the pure quantum regime, where
the WKB approximation used here is not valid, we infer that it could
be possible that going beyond the WKB approximation, would lead to
even higher frequencies. It could then lead to shorter switching
lengths than those displayed in
 Fig. (\ref{fig:fignuZ}-b)). In the
 semi-classical regime, the switching length $Z$ is
longer, nevertheless it is much shorter than the half period of
switching in the CW linear case, $L_{c}=\frac{\pi}{2\kappa^{phys}}$.
Indeed a pulse duration $\tau_{0}=0.6 ps$, and a silica fiber, one
has $L_{D}=16m$, that gives $L_{c}=125\frac{\kappa_{c}}{\kappa}$
when using $k_{c}=0.2$ (see Fig.(4)). For
$\frac{\kappa_{c}}{\kappa}=100$, the linear half period is
$L_{c}=12.5 km$ , which is
 several order of magnitudes longer than the semiclassical switching
 length $Z=3m$ (dashed curve).

\section{Summary and discussion}
\label{sec:discussion}

Even though the tunneling phenomenon is very familiar in many
wave-propagation problems, where the "true" wave-vector
$\overrightarrow{k}$ becomes $i\overrightarrow{k}$ after passing
under a classical barrier, (as in the case of evanescent waves in
the Fresnel theory), it appears in the present context in a slightly
unusual form: starting from the classical model (\ref{eq:coupled1}),
(\ref{eq:coupled2}) for the field enveloppe, which looks strangely
similar to the Schr\"odinger equation, our treatment based on the
approximate trial functions leads finally to the Euclidean system
(\ref{eq:4eqs}) which is not the Schr\"odinger equation for a wave
function.

 Within the trial functions approximation, the WKB or quasiclassical limit gave
 us the possibility
of estimating  rather easily the rate of {\emph{quantum}} tunneling
of a single soliton from one fiber to the other, even though it
should remain {\emph{classically}} in the same fiber forever. We
found that this rate of tunneling is not small and could well be
within reach of present day-experiments.

In the frame of the WKB approximation, we are trying to extend our
results by getting rid of the trial functions approximation. Our aim
is to check if the relation (\ref{eq:lnk}), that has been shown here
to survive when going from two to four unknown parameters in the
trial functions approximation, is valid beyond this approximation.
The calculation is heavier than the one presented here, because the
time $t$ is now considered as an infinite dimensional parameter,
then the semi-classical trajectory must be calculated from a set of
4 coupled PDE's, in place of the 4 coupled ODE's (\ref{eq:4eqs})
solved here. To derive these PDE's, we can choose for example
 $(\Im(u),\Im(v))$ and $(\Re(u),\Re(v))$  as set of conjugate variables
$(\textbf{p,q})$  for the classical system (\ref{eq:coupled1}),
(\ref{eq:coupled2}), with Hamiltonian

\begin{equation}
{\mathcal{H}_{NLS}} = \int \mathrm{d}t \left[
\frac{1}{2}|\frac{\partial u}{\partial t}|^2 - \frac{1}{2}|u|^4 +
\frac{1}{2} |\frac{\partial v}{\partial t}|^2 - \frac{1}{2}|v|^4 -
\kappa \left(u{\overline{v}}+ v{\overline{u}}\right)\right]
\mathrm{.} \label{eq:NLShamiltonian}
\end{equation}
 The Euclidean version of equations (\ref{eq:coupled1}),
(\ref{eq:coupled2}) is then obtained by changing $(z,\Im(u),\Im(v))$
into $(iz,i\Im(u),i\Im(v))$. The important point is that the
tunneling factor does not depend on the choice of $(\textbf{p,q})$,
while the Euclidean system obviously does. Finally, the heteroclinic
Euclidean trajectory is the  solution connecting two of the
classically permitted orbits, from $z=-\infty $ to $z=+\infty $. For
a given energy $E$, these are defined by the integro-differential
equation ${\mathcal{H}_{NLS}} = E$, where $\Im(u)=\Im(v)=0$,  and
correspond to the two asymmetrical solitons.

Because we have found that the Euclidean action $S_{E}^{Phys}$ can
be of order of $\hbar$ or even smaller in realistic experimental
conditions, it could even happen that the WKB quasiclassical
approximation is not valid anymore for computing the rate of
transfer from one fiber to the other. Usually the order of magnitude
of the action involved in the soliton picture, even in a single
fiber, is tacitly assumed to be far bigger than $\hbar$, which is an
assumption distinct from the one of a soliton made of many photons.
Indeed the soliton picture addresses perturbations to this `bound
state' of many photons that may be small enough to imply variations
of the action of order of $\hbar$, and so require some sort of
(`second') quantization. We plan to come to this general question in
future work, and outline here some of the estimated problems.

A treatment using the trial function, but valid beyond the WKB
approximation, is obviously more complicated than what we did here,
and perhaps questionable. Indeed it needs to consider both the trial
functions and their parameters as operators. Moreover it amounts to
assume that the fluctuations in $t$ and $z$ are decoupled, and, last
but not least, our result derived in the WKB approximation likely
signals that the assumption behind the classical (meaning non
quantum) theory for describing soliton in coupled fibers does not
hold anymore and that the quantum picture has to be used from the
start, which makes it theoretically challenging.

We assumed that every phenomenon under study involved solitons seen
as a coherent quantum objects. We argued that this requires that any
typical time, the tunneling time in  particular, is far shorter than
the coherence time. This coherence  time is of  order
of $t_c/N$ , with $t_c$ coherence time of a single photon in the
soliton, i.e. its mean-free flight time without change in phase or
 frequency. Because of the division by $N$ this may be a very short time. At times
longer than the coherence time any physical effect related to the
quantum coherence between states of solitons propagating in either
fiber is washed out. The final state, as described in the density
matrix formalism, is a state  of equal probability of the soliton on
either side without nondiagonal element. The experimental
manifestation of this state will be a  probability $1/2$ of
observing  a soliton in either fiber  without any possible
interference between
   the states on either side. Somehow this will bring the system back to  a fully
   classical state, except that this classical state has a probabilistic
   underpinning that is absent from the classical system:  in the   fully
    classical system the soliton  remains always in the same fiber, although
     in the quantum one its final state has a  probabilistic nature.

Looking at the other side of the coin one realizes that, because the
soliton is a composite object, and if it remains coherent during a
sufficiently long amount of time, its phase is the phase of a single
photon multiplied by the number of photons. Therefore any
interference experiment between coherent soliton states will have
much narrower interfringe than with a single photon or incoherent
photons, this interfringe being the one for a single photon divided
by the number of photons making the coherent soliton. This could be
of interest for gyroscopes based on the Sagnac effect \cite{Sagnac}.

Indeed a central issue concerning the observability of the tunneling
effect we present in this communication is the one of the quantum
coherence of the soliton, related itself to all dissipative effects
that can break up this coherence, and that makes the main topic
discussed in the present special issue. Nevertheless, even if the
coherence is limited, there is still tunneling, but at a reduced
rate \cite{PomSim}. In that case we suggest to use twin fibers with
coupling coefficient $\kappa$ periodically modulated in $z$, in
order to stimulate the switching process.

\begin{acknowledgments}
 Elisabeth Ressayre et Jean Ginibre  are gratefully acknowledged
for stimulating discussions, and Laurent Di Menza for providing us a
code for  simulations of the NLS model with transparent boundary
conditions.
\end{acknowledgments}

\begin{appendix}
\section{Physical units}
\label{appendix}

This appendix is about the relationship between quantities measured
in physical units for a standard fiber carrying solitons and the
dimensionless quantities used in the bulk of our paper. We
relate first the number of photons in a typical soliton to its time
duration, a duration called $\tau_{0}$ that we shall use afterwards
to give various order of magnitudes pertinent to our problem.

We use the standard expression of the electric field in a fiber,
written with the same notations as in the book by Newell and Moloney
\cite{newell}. The electric field of the EM wave in the fiber is
modelized by the wave-packet expression, $E=R(x_{t})B(z,t)\exp
(i\omega _{0}t-k_{0}z)+c.c.$ which obeys in a first approximation
 Maxwell's equations, when the duration of the pulse is not too
short, $x_{t}$ being the transverse coordinate, and $R(x_{t})$ the
dimensionless radial amplitude with  $\int \left| R(x^{\prime
})\right| ^{2}dx^{\prime }=\sigma $ as the core area.

Taking  $\frac{\delta \omega}{\omega_{0}}$ as a small parameter,
$\delta \omega$ being the frequency width of the pulse, one obtains the
NLS equation written in variables $z,\tau =t-z/v_{g}$, the
nonlinear and the dispersion term having
opposite signs:
\begin{equation}\label{eq:NLS}
i\frac{\partial B}{\partial z}+\frac{k_{0}^{"}}{2}\frac{\partial
^{2}B}{\partial\tau ^{2}}+\frac{2\pi n'_{2}}{\lambda _{0}}\left|
B\right| ^{2}B = 0 \mathrm{.}
\end{equation}
Note that we turned to the standard writing of the coefficients
of the NLS equation, $n'_{2}$ being the
modulus of the coefficient of the cubic Kerr effect and $k_{0}^{"}$
the modulus of the second derivative of the wavenumber $k_{0}$ with
respect to the the frequency of the EM wave.

The soliton solution is
\begin{equation}
B(z,t-z/v_{g})=B_{m}{\mathrm{sech}}\left((t-z/v_{g})/\tau
_{0}\right) \mathrm{,} \label{eq:sol}
\end{equation}

 \ with
\begin{equation}
\beta \left| B_{m}\right| ^{2}=1/\tau _{0}^{2} \label{eq:beta}
\mathrm{.}
\end{equation}

where $ \beta=\frac{2\pi n'_{2}}{\lambda _{0}k_{0}^{"}}\mathrm{.}$
Its energy is

\begin{equation}
W=\int P(t)dt=\sigma \int I(t){\mathrm{d}}t
\mathrm{,}
\end{equation}

where $I(t)$ is the optical intensity measured in watt per square
meter, and $P(t)$ is the Poynting vector integrated across the fiber
section,
 with a result expressed in Watts:
\begin{equation}
P(t)=\int (\mathbf{E\wedge H}). \mathbf{ z} dS
\end{equation}
where $\mathbf{z}$ is the unit vector in the direction of  propagation.

The magnetic field in the wave (supposing it is linearly polarized
with the electric field in the $x$-direction) is $H_{x}=n
c\varepsilon _{0}E_{y}$, n index of refraction. For a  material with
instantaneous response, after
averaging over one period of the field oscillations, one finds

\begin{equation}\label{eq:poynting}
\overline{P}(t)=2c\varepsilon _{0}\sigma n\left| B(t)\right| ^{2}
\mathrm{.}
\end{equation}

At leading order, i.e. by taking into account the linear part of
refractive index, $n=n_{0}$, this yields
\begin{equation}
I^{(0)}(t)=2c\varepsilon _{0}\ n_{0}\left| B(t)\right|^{2}
\mathrm{.}
\end{equation}
Whence the energy of the pulse is:

\begin{equation}
W^{(0)}=N\hbar \omega =2n_{0}c\varepsilon _{0}\sigma \int \left|
B(t-z/v_{g})\right| ^{2}dt=4n_{0}c\varepsilon _{0}\sigma
\frac{\lambda _{0}k_{0}^{"}}{2\pi n'_{2}}\frac{1}{\tau _{0}}
\mathrm{.}
\end{equation}

Finally the relationship between the photon number and the pulse
duration  writes

\begin{equation}
N \tau _{0} =4n_{0}\varepsilon _{0}\sigma (\frac{\lambda
_{0}}{2\pi})^{2}\frac{k_{0}^{"}}{n'_{2}\hbar} \mathrm{.}
\end{equation}
Similarly the action is
\begin{equation}
S^{(0)}=N\hbar =W^{(0)}/\omega \mathrm{,}
\end{equation}
 at leading order.
\bigskip

In MKSA units, with standard values (see \cite{newell}) of optical
fibers composed of silica cores, this gives:

$n_{0}=1.5$

$\epsilon _{0}=0.89\mathrm{.}10^{-11}F/m$, or $\epsilon
_{0}c=1/Z_{0}$ with $Z_{0}=377 \Omega$ the impedance of free space,

$\hbar =10^{-34}J.s$

$\lambda _{0}=1.55\mathrm{.}10^{-6}m$

$\sigma \sim 10^{-11}$ for a $10\mu m ^{2}$-area fiber.

$k_{0}^{"}=2.2\mathrm{.}10^{-26}s^{2}/m$

$n'_{2}= 2.6\mathrm{.}10^{-22}(m/V)^{2}$

 With  these data, the number of photons in the pulse of duration
$\tau_{0}$, measured in seconds obeys the relation:

\begin{equation}
N \tau _{0}\sim  3\mathrm{.}10^{-5}\mathrm{,}
\end{equation}
that gives  $N\sim 3\mathrm{.}10^{7}$ photons for a  $ps$-pulse.

 Let
us note that the nonlinear index of refraction $n'_{2}$ may be
several orders of magnitude larger, when using other materials. For
example, in the experiment of Wa et al. \cite{wa}, the optical
switch was studied in multiple quantum well wave-guides, with
$n'_{2}=10^{-13} (m/V)^{2}$.  \\

\bigskip

\textbf{Coherent part of the energy and action}\\

\bigskip

Let us write the energy and action as
\begin{equation}
\mathcal{W}=W^{(0)}+W^{(1)} \mathrm{,}
\end{equation}
and
\begin{equation}
\mathcal{S}=S^{(0)}+S^{(1)} \mathrm{.}
\end{equation}

The dominant contributions are proportional to the linear part of
the refractive index, namely a term contained in the Maxwell
equation. The subdominant contributions $W^{(1)}$ and $S^{(1)}$ ,
correspond to the terms contained in the envelope equation, they are
perturbations to the dominant effects calculated above. For two
coupled fibers, these perturbations result from balanced effects of
dispersion, coupling and nonlinearity. They are proportional to the
energy
 and action of the the
dimensionless NLS equation (\ref{eq:coupled1}-\ref{eq:coupled2}),

\begin{equation}
\left \{ \begin{array}{l}

W^{(1)}=w^{(1)}{\mathcal{H}}_{\mathrm{NLS}}\\
S^{(1)}=\frac{w^{(1)}}{\omega_{0}}{\mathcal{S}}_{\mathrm{NLS}}
\end{array}
\right. \label{eq:w1}
\end{equation}

where the scaled energy ${\mathcal{H}}_{\mathrm{NLS}}$ is defined in
equation (\ref{eq:NLShamiltonian}) , and the action in equation
(\ref{eq:NLSLagrdimcoupl}).

 The coefficient $w^{(1)}$ may be calculated by using the expression of
the Poynting vector (\ref{eq:poynting}) valid for dispersionless
Kerr media, where
\begin{equation}
n=n_{0}+n_{2}I =n_{0}+n'_{2}\left| B(t)\right| ^{2} \mathrm{,}
\end{equation}

This gives \ $W^{(1,Kerr)}=2n'_{2} c\epsilon _{0}\sigma \int
dt\left| B(t)\right| ^{4}$
.

Taking the hyperbolic secant solution (\ref{eq:sol}-\ref{eq:beta}),
 one obtains $ W^{(1,Kerr)}=2n'_{2} c\epsilon _{0}\sigma \left|
B_{M}(t)\right| ^{4}\tau_{0}\int dt sech^{4}(t)$, or

\bigskip
\begin{equation}
\ W^{(1,Kerr)}= -4 n'_{2} c\epsilon _{0}\sigma
\frac{1}{\beta^{2}\tau_{0}^{3}}{\mathcal{H}}^{\mathrm{Kerr}}_{\mathrm{NLS}}
\mathrm{,} \label{eq:w13}
\end{equation}
where ${\mathcal{H}}^{\mathrm{Kerr}}_{\mathrm{NLS}}$ is the Kerr
contribution of the Hamiltonian (second term in the r.h.s. of
equation \ref{eq:energie}). This correction is the "coherent" part
of the energy in the sense that it is proportional to the square of
the intensity, or of the photon number. Finally we are ready to
express  the physical value of the Hamiltonian and action
 associated to a soliton whose temporal width is scaled to
$\tau_{0}$ as it was assumed in sections 3-4, by using the
expression (\ref{eq:w1}) with

\begin{equation}
w^{(1)}\simeq 4 n'_{2} c\epsilon _{0}\sigma
\frac{1}{\beta^{2}\tau_{0}^{3}} \label{eq:w14} \mathrm{.}
\end{equation}

 This allows in particular to
express concretely the constraint that nonlinear effects are small,
that is that $W^{(1)}<<W^{(0)}\mathrm{,}$ a condition equivalent to

\begin{equation}
\frac{W^{(1)}}{W^{(0)}}=\frac{2}{3}\frac{n'_{2}}{n_  {0}}%
\frac{1}{\beta \tau _{0}^{2}}<<1 \mathrm{.}
\end{equation}

Finally the physical action  associated to the quasi-classical
trajectory is approximately given by the expression
\begin{equation}
\mathcal{S}_{E}^{phys}=\frac{w^{(1)}}{\omega_{0}}S_{E} \mathrm{.}
\end{equation}

The quantum tunneling coefficient $T=exp{(-2{S}_{E}^{phys}/\hbar})$
is expected to be experimentally observable when the physical action
is not too large with respect to $\hbar$. When $S_{E}^{phys}$
becomes of order $\hbar$, we can infer that we are in the "pure
quantum" regime.  The frontier between these two regimes may be
drawn in the space parameters ($\kappa, \tau_{0}$) by introducing
the parameter

\begin{equation}
\gamma=\frac{8 c^{3}\epsilon _{0}\sigma k"_{0}^{2}}{\hbar n'_{2}}
\label{eq:gamma}
\end{equation}
that allows to write
\begin{equation}
2\frac{w^{(1)}}{\hbar
\omega_{0}}=\frac{\gamma}{(\omega_{0}\tau_{0})^{3}} \mathrm{.}
\end{equation}

With the data given above for standard fibers, $\gamma=3.6 10^{8}$,
the semi-classical regime stands below the curve drawn in section
\ref{sec:quantization} (see Fig. (5)). Consequently pulses longer
than one $ps$ typically stands in the pure quantum regime. In
conclusion we predict quantum tunneling for realistic conditions of
soliton propagation in two coupled fibers.

\end{appendix}

{}

\ifx\mainismaster\UnDef%

\thebibliography{9}

\bibitem{faddeev} L. D. Fadeev, V. E. Korepin, Phys. Rep. C {\textbf {42}}, (1978) 270.

\bibitem{Feynman} R. P. Feynman , R. B. Leighton and M. Sands, \emph  {Lectures on
Physics}, (Add-Wesley Publ., Readings, 1964).

\bibitem{PomSim}Y.
Pomeau, A. Pumir, J. de Phys. (Paris) {\bf{46}}, 1797 (1985).

\bibitem{Akh} N. Akhmediev, A. Ankiewicz,
Phys. Rev. Lett. {\bf{70}}, 2395 (1993); J. M. Soto-Crespo and N.
Akhmediev, Phys. Rev. E {\bf{ 48}}, 4710 (1993);  N. Akhmediev, J.
M. Soto-Crespo Phys. Rev. E {\bf{49}}, 4519 (1994); A.V. Buryak, N.
N. Akhmediev, IEEE Journal of Q.Elect. {\bf{31}}, 682 (1995).


\bibitem{griffiths} D.J. Griffiths, \emph{Quantum
Mechanics} (Prentice Hall, 2004); see also L. D. Landau and E. M.
Lifchitz in \emph{M\'ecanique Quantique, Th\'eorie non relativiste},
(Ed. Mir, Moscou ,1966).

\bibitem{Malomed} B. A. Malomed, I. M. Skinner, P. L. Chu and
G. D. Peng, , Phys. Rev. E {\bf{ 53}}, 4084 (1996); I. M. Uzunov, R.
Mushall, M. Golles, Yu. S. Kivshar, B. A. Malomed, and F. Lederer,
Phys. Rev. E {\bf{51}}, 2527 (1995).

\bibitem{Trillo} S. Trillo, S. Wabnitz, E. M. Wright and G.I.
Stegeman, Opt. Comm. {\bf{70}}, 166 (1989); E. M. Wright, G.I.
Stegeman and S. Wabnitz, Phys. Rev. A {\bf{ 40}}, 4455 (1989); S.
Trillo, S. Wabnitz, E. M. Wright and G.I. Stegeman, Opt. Lett. A
{\bf{ 13}}, 871 (1988); and Opt. Lett. A{\bf{13}}, 672 (1988)

\bibitem{Pare} C. Par\'e, M. Florjanczyk, Phys.Rev. A {\bf{
41}}, 6287 (1990).


\bibitem{Kivshar}Y.S. Kivshar, Opt. Lett. {\bf{18}}, 7 (1993).

\bibitem{smyth} N. F. Smyth, A. L. Worthy, J. Opt. Soc. Am. {\bf{
14}}, 2610 (1997); N. F. Smyth , A. H. Pincombe, Phys. Rev. E {\bf{
57}}, 7231 (1998).

\bibitem{pomrotation}
Y. Pomeau, Europhys. Lett., {\bf{74}} 951 (2006).

\bibitem{pomenonlin} Y. Pomeau, Nonlinearity {\bf{5}}, 707 (1992).

\bibitem{landau} L. D. Landau and E. M. Lifchitz in \emph{Mechanics   and Electrodynamics}
(Pergamon Press, N.Y., 1972).

\bibitem{coleman}
S.Coleman, in \emph{Proc. Int.  School of Subnuclear Physics},
Europhys. (Erice , 2006); and in \emph{Aspects of
symmetry}(Cambridge University Press, 1985).


\bibitem{newell}  A. C. Newell and J.V. Moloney \emph{Nonlinear Optics},
( Addison-Wesley Publ. Company, 1992).
\bibitem{erlangen} A. Sizmann, Appl. Phys. B {\bf{65}}, 745 (1997).

\bibitem{Sagnac} E. J. Post, Rev. of Mod. Phys.
{\bf{39}}, 475 (1967).

\bibitem{wa} P. Li Kam Wa, J.E. Sitch, N. J. Mason, J. S. Roberts, P. N. Robson, Electronics Letters,
{\bf{21}}, 27 (1985)


\endthebibliography
\end{document}